\documentstyle[11pt,cite,axodraw]{article}

\newcommand{\be}{\begin{equation}}
\newcommand{\ee}{\end{equation}}
\newcommand{\bea}{\begin{eqnarray}}
\newcommand{\eea}{\end{eqnarray}}

\newcommand{\nn}{\nonumber}

\newcommand{\tA}{\widetilde{A}}

\newcommand{\eq}[1]{Eq.~(\ref{#1})}

\newcommand{\NPB}[3]{Nucl.\ Phys.\ {\bf B{#1}} (19{#2}) {#3}}
\newcommand{\PRD}[3]{Phys.\ Rev.\ {\bf D{#1}} (19{#2}) {#3}}
\newcommand{\PLB}[3]{Phys.\ Lett.\ {\bf B{#1}} (19{#2}) {#3}}
\newcommand{\PRL}[3]{Phys.\ Rev.\ Lett.\ {\bf {#1}} (19{#2}) {#3}}

\newcommand{\ZPC}[3]{Z.\ Phys.\ {\bf C{#1}} (19{#2}) {#3}}
\newcommand{\EPJC}[3]{Eur.\ Phys.\ J.\ {\bf C{#1}} (19{#2}) {#3}}

\newcommand{\ks}{{\mbox k \!\!\! /}}
\newcommand{\ps}{{\mbox p \!\!\! /}}

\newcommand{\im}{{\rm Im\,}}
\newcommand{\re}{{\rm Re\,}}

\begin{document}
 
\begin{center}
 
{\Large 
{\bf The Pinch Technique Approach To}
\\[0.1cm]
{\bf Gauge-Independent $n$-Point Functions}
}

\vspace*{0.4cm}

{\bf N.J. Watson} 
\\
Rutherford Appleton Laboratory,
Chilton, Didcot,
Oxfordshire OX11 0QX,
UK. 
\\
jay.watson@rl.ac.uk 

\end{center}

\begin{abstract}

The pinch technique (PT) is an algorithm for the rearrangement of
contributions to conventional, gauge-dependent $n$-point functions in
gauge theories to obtain a formulation of one-loop perturbation
theory in terms of ``effective'' $n$-point functions which, principle
among many other desirable properties, are individually entirely
independent of the particular gauge-fixing procedure used. The aim
of this talk is to give an introductory account of
(i) the phenomenological motivations for the PT approach, 
(ii) the PT algorithm, including the principle properties of
the PT $n$-point functions, (iii) an example application, and 
(iv) recent progress in extending the PT beyond the one-loop level.

\end{abstract}
 

\vspace*{0.3cm}

{\large {\bf 1. Motivation}}

\noindent 
In order to quantize a gauge theory, it is necessary to remove
the redundant degrees of freedom resulting fron the gauge symmetry.
The standard procedure thus
involves adding to the classical lagrangian
a gauge-fixing term, together with an
associated Fadeev-Popov ghost term:
\be
{\cal L}(\mbox{classical})
\,\longrightarrow\,
{\cal L}(\mbox{classical})
+{\cal L}(\mbox{gauge-fixing})
+{\cal L}(\mbox{ghost})\,\,.
\label{L}
\ee
The gauge symmetry of the classical lagrangian is
thus replaced by the BRST symmetry of the gauge-fixed lagrangian.
This procedure has two immediate consequences.
First, the individual $n$-point (Green's) functions in general
depend explicitly and non-trivially on the particular choice
of the gauge-fixing term in (\ref{L}). 
Second, for a non-abelian theory,
the ensemble of $n$-point functions satisfy complicated Slavnov-Taylor
identities involving ghost fields and
associated with the BRST invariance of the gauge-fixed lagrangian,
rather than the gauge invariance of the original classical lagrangian.

As long as one is interested in computing strictly order-by-order
in perturbation theory
$S$-matrix elements for the scattering of on-shell fields,
neither of the above two facts presents a problem.
In particular, order-by-order the $S$-matrix element 
(assumed to be defined) for a given process 
is guaranteed by general proofs to be independent of the given
gauge-fixing procedure. Thus, one is at liberty to choose
the gauge (assumed to be renormalizable etc.)
in which the calculations are simplest, safe in the
knowledge that the final result, to any order in perturbation
theory, must be independent of this choice. 

However, there are (at least) two classes of 
application for which the
conventional approach to perturbation theory just outlined
is not sufficient:

\begin{enumerate} 

\item
{\em Partial summations of perturbation theory.}
There are various situations in which one wishes to sum up
some infinite subset of contributions to a given
process, i.e.\ go beyond order-by-order computation. 
The simplest example of this is the Dyson summation
of the self-energy of a given field. Such a summation 
occurs in two closely-related cases:

\begin{itemize}

\item
{\em Unstable particles.}
In order to regulate the pole which occurs in the
tree level propagator for such particles, one is obliged to Dyson-sum the
self-energy so that it appears in the
denominator of the radiatively-corrected propagator.
The imaginary part of the self-energy then specifies
the decay width of the given particle.

\item
{\em Effective charges}. 
In order to provide a well-defined basis for renormalon analyses,
and also to improve predictions in phenomenology,
one would like to be able to extend 
directly at the diagrammatic level the 
Gell-Mann--Low concept of an effective charge 
from QED to non-abelian gauge theories.

\end{itemize}

In both cases, the fundamental problem is the ambiguity of the 
self-energies involved due to their gauge dependence:
in general, the self-energies are gauge-dependent
at all $q^{2}$ away from the propagator pole position.
Furthermore, even if this gauge dependence is ``mild''
(usually a meaningless concept), the naive Dyson summation
in general leads to potentially catastrophic
violations of gauge invariance (i.e.\ current conservation)
when embedded in a matrix element.

\item
{\em Explicitly off-shell processes.}
The second class of applications occurs when one wishes to
consider amplitudes for processes in which some or all
of the external fields are off-shell. Such cases include:

\begin{itemize}

\item
{\em Form factors.} A popular way of comparing theory with
experiment is via the use of form factors
to parameterize loop corrections to vertices
involving off-shell fields,
e.g.\ the $\gamma W^{+}W^{-}$ and $ZW^{+}W^{-}$ vertices
measured at LEP. 
Similarly, electroweak ``oblique'' i.e.\ self-energy corrections are
often parameterized in terms of $S,T,U$.

\item
{\em Matching of full and effective gauge theories.}
The description of the low-energy physics of a theory
involving light and heavy fields via an effective lagrangian
requires a matching at low energies of the off-shell
$n$-point functions of the full and effective theories.

\item
{\em Schwinger-Dyson analyses.} While in principle
able to give exact, non-perturbative information,
in practice the Schwinger-Dyson equations for gauge theories 
require some truncation in order to be tractable,
together with the use of 
perturbative $n$-point functions as building blocks.

\end{itemize}

Again, in all of these examples, a
fundamental problem is the ambiguity of the
$n$-point functions involved due to their gauge dependence.
At best, in the above cases,
this ambiguity necessitates very careful consideration
when comparing theory with experiment
(e.g.\ when looking for signals of ``new physics''); 
at worst, it renders analyses useless.

\end{enumerate}

The PT \cite{cornwall0,cornwall1,cornpapa1,kennlynn,papa1,degrsirl,papaphil,
papaparr,jay1,PT2pt,PTqcdeffch,PTeweffch,higgs,PT2loop,QCD98}
provides a framework for perturbation theory
in which these problems are tackled at source.
In particular, the $n$-point functions obtained in the
PT approach are {\em gauge-independent}, i.e.\ 
they are entirely independent of
the particular gauge-fixing procedure used
(independent of $\xi$, $n_{\mu}$ etc.), and also
{\em gauge-invariant}, i.e.\ they satisfy simple
tree-level-like Ward identities associated with the
gauge symmetry of the classical lagrangian of the theory. 
Note that, throughout, we distinguish between gauge independence
and gauge invariance; the former does {\em not}\, follow from the latter.
It is important to emphazise that, in addition to being both
gauge-independent and gauge-invariant,
the PT $n$-point functions display a wide variety of
further desirable theoretical properties.

\vspace*{0.3cm}

{\large {\bf 2. The Pinch Technique Algorithm}}

\noindent
The PT is based on the observation \cite{cornwall0} that 
one-loop Feynman diagrams that naively contribute
to a given $n$-point function in fact in general 
contain components identical in structure to diagrams
which contribute to $m$-point functions, $m < n$.
In order to see this, consider the now-canonical case of
quark-quark scattering at one loop in QCD \cite{cornpapa1}. 
The corrections to the 

\begin{center}
\begin{picture}(425,95)(73,340)


\ArrowLine(100,360)(100,390)
\ArrowLine(100,390)(100,420)
\Gluon(100,390)(140,390){2.5}{6}
\Gluon(140,390)(180,410){2.5}{6.5}
\Gluon(140,390)(180,370){-2.5}{6.5}
\ArrowLine(180,360)(180,420)
\put(100,350){\makebox(0,0)[c]{\small $p^{\prime}$}}
\put(100,430){\makebox(0,0)[c]{\small $p^{\prime}-q$}}
\put(180,350){\makebox(0,0)[c]{\small $p$}}
\put(180,430){\makebox(0,0)[c]{\small $p+q$}}
\put(140,340){\makebox(0,0)[c]{\small (a)}}

\put(200,390){\makebox(0,0)[c]{\Large $=$}}

\ArrowLine(220,360)(220,390)
\ArrowLine(220,390)(220,420)
\Gluon(220,390)(262.5,390){2.5}{6}
\GlueArc(280,390)(17.5,180,540){2.5}{16}
\ArrowLine(300,360)(300,390)
\ArrowLine(300,390)(300,420)
\put(260,340){\makebox(0,0)[c]{\small (b)}}

\put(320,390){\makebox(0,0)[c]{\Large $+$}}

\ArrowLine(340,360)(340,390)
\ArrowLine(340,390)(340,420)
\Gluon(340,390)(380,390){2.5}{6}
\Gluon(380,390)(420,410){2.5}{6.5}
\Gluon(380,390)(420,370){-2.5}{6.5}
\Line(395,390)(410,397.5)
\Line(395,390)(410,382.5)
\ArrowLine(420,360)(420,420)
\put(380,340){\makebox(0,0)[c]{\small (c)}}
\put(375,405){\makebox(0,0)[c]{$\Gamma^{F}$}}
\put(400,365){\makebox(0,0)[c]{$D^{F}$}}
\put(400,415){\makebox(0,0)[c]{$D^{F}$}}

\put(437,390){\makebox(0,0)[l]{\Large $+ \,\,\cdots$}}

\end{picture}
\end{center}

\noindent{\small
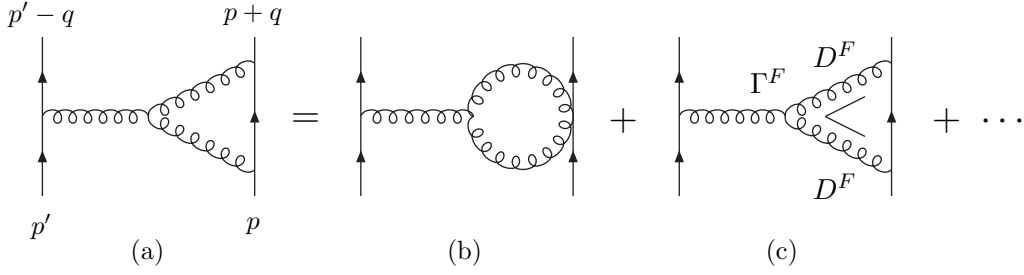
Fig.\ 1. A conventional vertex diagram (a), its
self-energy-like ``pinch part'' (b) and ``non-pinch part'' (c).
The ``wedge'' in (c) indicates and
specifies the orientation of the vertex component 
$\Gamma^{F}$ defined in \eq{tgv}.
The ellipsis represents terms which vanish when
the external fermions are on-shell. }

\vspace{0.3cm}

\noindent
tree level amplitude involve self-energy (2-point),
vertex (3-point) and box (4-point) diagrams.
In the class of linear covariant gauges with gauge parameter $\xi$
($\xi = 0$ is the Landau gauge), the conventional 
one-loop gluon self-energy (2-point function) is given by
\be
i\Pi_{\mu\nu}^{(1)}(q,\xi) 
= 
i(q^{2}g_{\mu\nu} - q_{\mu}q_{\nu})\Pi^{(1)}(q^{2},\xi)\,\,,
\ee
with, for $n_{f}$ flavours of massless quark,
in $d=4-2\epsilon$ dimensions,
\bea
\Pi^{(1)}(q^{2},\xi)
&=&
\frac{g^{2}}{16\pi^{2}}\biggl\{ 
\biggl[ \frac{3\xi -13}{6}C_{A} + \frac{4}{3}T_{F}n_{f}\biggr]
\biggl[ - C_{\rm UV} + \ln\biggl(\frac{-q^{2}}{\mu^{2}}\biggr)\biggr]
\nn \\
& &
+ \,\frac{9\xi^{2} + 18\xi + 97}{36}C_{A} - \frac{20}{9}T_{F}n_{f}\biggr\}
\,\,-\,\,(Z_{3}-1)^{(1)}\,\,,
\eea
where $\mu$ is the 't~Hooft mass scale and
$C_{\rm UV} = \epsilon^{-1} + \ln(4\pi) - \gamma_{E}$.
However, the vertex and box diagrams also
have components---the so-called ``pinch parts''---which 
have exactly the same structure as the gluon
self-energy correction.
The identification of these pinch parts involves two steps
\cite{cornwall0,cornwall1,cornpapa1}: first, the isolation 
in the {\em integrands}\, for the diagrams of all
factors of longitudinal four-momentum associated with the gauge fields
propagating in the loops; 
and second, the systematic use of
the elementary Ward identities obeyed by
the tree level vertices when contracted with these longitudinal factors.

In order to see how this works, consider the
one-loop vertex diagram shown in Fig.~1a. 
The contribution of this diagram is given 
in standard notation by
\bea
\mbox{Fig.\,1a}
&=&
\mu^{2\epsilon}\int\frac{d^{d}k}{(2\pi)^{d}}\,
ig\gamma_{\mu'}T^{m}\,iD^{\mu'\!\mu}(q)
\overbrace{
gf^{mrs\phantom{'}}\!\Gamma_{\mu\rho\sigma}(q,k_{1},k_{2})
}^{\mbox{\scriptsize triple gauge vertex}}
\overbrace{
iD^{\rho\rho'}(k_{1})\,iD^{\sigma\sigma'}(k_{2})
}^{\mbox{\scriptsize internal gauge propagators}}
 \nn \\
& & \times\,
ig\gamma_{\sigma'}T^{s}\,iS(p-k_{1})\,ig\gamma_{\rho'}T^{r}\,\,,
\label{fig1a}
\eea
where $S(p-k_{1}) = (\ps - \ks_{1} -m + i\epsilon)^{-1}$ is the
internal fermion propagator (here $m=0$).
In \eq{fig1a}, the sources of longitudinal internal gauge field
four-momentum $k_{1\rho'}$, $k_{2\sigma'}$ are
the two internal gauge propagators and the triple gauge vertex.
For the latter, one writes \cite{cornwall0,cornwall1,cornpapa1}
\be
\Gamma_{\mu\rho\sigma}(q,k_{1},k_{2})
=
\underbrace{
\overbrace{
(k_{1}-\phantom{\Bigl|}\!\!k_{2})_{\mu}g_{\rho\sigma} 
}^{\mbox{\scriptsize convection}}
\,\,\,
\overbrace{
-\phantom{\Bigl|}\!2q_{\rho}g_{\sigma\mu}+2q_{\sigma}g_{\rho\mu}\phantom{|}\!
}^{\mbox{\scriptsize spin-1 magnetic}} 
}_{\displaystyle \Gamma_{\mu\rho\sigma}^{F}(q;k_{1},k_{2})}
\,\,\,
\underbrace{
\overbrace{
-\phantom{\Bigl|}\!
k_{1\rho}g_{\sigma\mu} + k_{2\sigma}g_{\rho\mu}\phantom{|}\!
}^{\mbox{\scriptsize longitudinal}}
}_{\displaystyle \Gamma_{\mu\rho\sigma}^{P}(q;k_{1},k_{2})}.
\label{tgv}
\ee
The component $\Gamma_{\mu\rho\sigma}^{F}$
contributes no factors of longitudinal
internal four-momentum $k_{1\rho}$, $k_{2\sigma}$ and obeys
a Ward identity involving the difference of
inverse Feynman gauge propagators:
\be
q^{\mu}\Gamma_{\mu\rho\sigma}^{F}(q;k_{1},k_{2})
=
D_{\rho\sigma}^{F}(k_{1})^{-1} 
-D_{\rho\sigma}^{F}(k_{2})^{-1} \,\,.
\ee
When the longitudinal factors 
from the internal gauge propagators and the triple gauge vertex 
in \eq{fig1a} are contracted with the corresponding Dirac matrices 
occurring in the tree level gauge-fermion vertices,
they trigger the elementary Ward identities
\be
\ks_{1} 
=
S^{-1}(p) - S^{-1}(p-k_{1})\,,\qquad\qquad
\ks_{2}
=
S^{-1}(p-k_{1}) - S^{-1}(p+q)\,\,.
\label{wids}
\ee
The inverse fermion propagators $S^{-1}(p-k_{1})$ 
cancel (``pinch'') the internal fermion propagator
in \eq{fig1a}, leading to the pinch part of the
diagram shown schematically in Fig.~1b. Using 
$f^{mrs}T^{s}T^{r} = -\frac{1}{2}iT^{m}$, we see that
{\em the pinch part of the vertex (3-point) diagram
couples to the pair of external fermion
lines in exactly the same way as the conventional
gluon self-energy (2-point) diagram.}
The terms in \eq{fig1a} involving no longitudinal factors,
i.e.\ the Feynman gauge component $D^{F}$
of the gauge propagators and the component 
$\Gamma^{F}$ of the triple gauge vertex, 
give the ``non-pinch'' or
``genuine'' vertex contribution shown in Fig.~2c.
The terms proportional to $S^{-1}(p)$ and $S^{-1}(p+q)$
in the Ward identities Eqs.~(\ref{wids}) give the
contribution denoted by the ellipsis in Fig.~1. This last
contribution vanishes when the external quarks are on-shell.

A similar procedure may be carried out for the remaining 
one-loop vertex and box diagrams for the process
in order to isolate their self-energy-like pinch parts.
Adding these pinch parts to the conventional self-energy,
one obtains the PT gluon ``effective'' 2-point function
\cite{cornwall0,cornwall1,cornpapa1}:
\be
\hat{\Pi}^{(1)}(q^{2})
=
\frac{g^{2}}{16\pi^{2}}\biggl\{ \beta_{0}
\biggl[ - C_{\rm UV} + \ln\biggl(\frac{-q^{2}}{\mu^{2}}\biggr)\biggr]
\,+\,\frac{67}{9}C_{A} - \frac{20}{9}T_{F}n_{f}\biggr\}
\,\,-\,\,(\hat{Z}_{3}-1)^{(1)}\,\,,
\ee
where $\beta_{0} = -\frac{11}{3}C_{A} + \frac{4}{3}T_{F}n_{f}$ is
the first coefficient of the $\beta$ function.
It is seen from the above expression that the PT self-energy
is entirely gauge-independent, with the coefficient of the logarithm
given by the $\beta$ function, just like the vacuum polarization
in QED.

The PT ``effective'' $n$-point functions,
$n= 3,4,\ldots$, i.e.\ vertices, boxes etc.,
may be similarly constructed 
\cite{cornpapa1,kennlynn,papa1,degrsirl,papaphil,papaparr}.
For example, the PT one-loop triple
gauge vertex (3-point) function 
$\hat{\Gamma}_{\alpha\beta\gamma}^{(1)}(q_{1},q_{2},q_{3})$
may be obtained from the consideration
of diagrams with three external quark lines.
It is found \cite{cornpapa1}
that this function is (i) gauge-independent, and (ii)
obeys a simple tree-level-like Ward identity associated with the gauge
invariance of the classical lagrangian:
\be
q_{1}^{\alpha}\hat{\Gamma}_{\alpha\beta\gamma}^{(1)}(q_{1},q_{2},q_{3})
=
\hat{\Pi}_{\beta\gamma}^{(1)}(q_{2})
-\hat{\Pi}_{\beta\gamma}^{(1)}(q_{3})\,\,,
\label{tgvwid}
\ee
where 
$\hat{\Pi}_{\beta\gamma}^{(1)}(q)
= (q^{2}g_{\beta\gamma} - q_{\beta}q_{\gamma})\hat{\Pi}^{(1)}(q^{2})$.
It is emphasized that, 
in the conventional approach in ordinary covariant gauges, 
the above Ward identity holds for the fermionic contributions but
not for the bosonic contributions.

The above cases are the simplest examples of the implementation
of the PT algorithm. In general, {\em the PT
is a systematic rearrangement of contributions 
to conventional one-loop $n$-point functions according simply
to the structure of their couplings to external fields.} 
This rearrangement takes place at the level of the Feynman integrands, 
and results in the set of
PT one-loop ``effective'' $n$-point functions for the given theory, 
be it unbroken, e.g.\ QCD, or spontaneously broken,
e.g.\ the electroweak standard model.
The principle properties of the
PT one-loop $n$-point functions may be summarised as follows:

\begin{itemize}

\item
{\em Gauge independence.} The PT one-loop $n$-point functions are
entirely independent of the particular gauge-fixing procedure used,
be it covariant or non-covariant, linear or non-linear
\cite{cornwall0,cornwall1,cornpapa1,kennlynn,papa1,degrsirl,papaphil,
papaparr,jay1,PT2pt,PTqcdeffch,PTeweffch,higgs}.

\item
{\em Gauge invariance.} The PT one-loop $n$-point functions satisfy
simple tree-level-like Ward identities associated with the gauge
invariance of the classical lagrangian
\cite{cornwall0,cornwall1,cornpapa1,kennlynn,papa1,degrsirl,papaphil,papaparr}.

\item
{\em Renormalizability.} The PT one-loop $n$-point functions are 
multiplicatively renormalizable by local counterterms, regardless
of the renormalizability of the initial gauge
\cite{cornwall0,cornwall1,cornpapa1,kennlynn,papa1,degrsirl,papaphil,papaparr}.

\item
{\em Universality.} The PT one-loop $n$-point functions are independent
of the quantum numbers of the external fields in the particular
process from which they are obtained \cite{jay1,PTqcdeffch}.

\item
{\em Analyticity and unitarity.} The PT one-loop $n$-point functions
display properties of analyticity and unitarity exactly analogous to
those of scalar field theories \cite{PT2pt,PTeweffch}.

\item
{\em Relation to observables.} The PT one-loop $n$-point functions are
in principle directly related to physical observables via dispersion
relations \cite{PTeweffch}.

\item
{\em Feynman rules.} The PT one-loop $n$-point functions for any 
given gauge theory may in fact be obtained directly using a 
well-defined set of Feynman rules \cite{BFMPT1,BFMPT2,BFMPT3}.

\end{itemize}

In addition to these general properties, specific PT $n$-point functions
display further important properties. In particular, it has been shown
that the PT 2-point functions may be summed in Dyson series
\cite{PT2pt,PTqcdeffch}, and that they
do not shift the position of the propagator pole \cite{PT2pt}.
It is the above collection of properties which has led to the
PT being advocated as the appropriate theoretical framework for
the wide range of applications described in the introduction for which
the conventional perturbative approach is inadequate 
\cite{PT2pt,PTqcdeffch,PTeweffch,higgs,PTapps}

\vspace*{0.3cm}

{\large {\bf 3. Relation to the Background Field Method}}

\noindent
It was observed some time ago \cite{BFMPT1,BFMPT2,BFMPT3} 
that the PT one-loop
$n$-point functions {\em coincide}\, with the background field
$n$-point functions obtained in the background field method (BFM)
in the Feynman-like quantum gauge $\xi_{Q} = 1$. This correspondence
holds for the cases both of unbroken and spontaneously broken
gauge symmetry. To obtain an heuristic explanation of 
this correspondence \cite{BFMPT1,BFMPT2,BFMPT3,PTquark}, 
consider the simpler unbroken case, e.g.\ QCD again. 

The basic idea of the BFM \cite{abbott} is first to make a shift
of the gauge field variable in the classical lagrangian:
$A \rightarrow \tA + A$, 
where $\tA$ is the background gauge field
and $A$ is the quantum gauge field. The quantum field
is the functional integration variable.
The gauge-fixing term for the quantum field
is then chosen so as to retain invariance of the generating functional
under gauge transformations of the background field:
\be
{\cal L}(\mbox{quantum gauge-fixing}) 
=
-\, \frac{1}{2\xi_{Q}}\Bigr[D_{\mu}^{ab}(\tA)A^{b\mu}\Bigr]^{2}\,\,,
\label{BFMgaugefixing}
\ee
where $D_{\mu}^{ab}(\tA)$ is the background covariant derivative.
The gauge-fixing term for the
background gauge field may then be chosen entirely independently. 
Demanding that the shift be made such that
the quantum gauge field has vanishing vacuum expectation value,
the background gauge fields are the classical fields which do not
propagate in loops (since the path integral is only over
the quantum fields), while the quantum gauge fields do not appear
as external fields (since $\langle A_{\mu}^{a} \rangle = 0$).

As a result of the explicitly-retained background gauge invariance,
the background field $n$-point functions are gauge-invariant,
i.e.\ they satisfy to all orders simple tree-level-like Ward identitites
associated with the gauge invariance of the classical lagrangian.
However, the background field $n$-point functions 
remain gauge-{\em dependent}, i.e.\ they still
depend explicitly on the gauge-fixing parameter $\xi_{Q}$.

At the level of Feynman rules, 
the effect of the non-linear gauge-fixing \eq{BFMgaugefixing}
is to alter not only the quadratic quantum gauge field terms
$AA$ from the classical lagrangian, but also the
cubic and quartic terms $\tA AA$ and $\tA\tA AA$ \cite{abbott}.
In particular, the Feynman rule for the $\tA AA$ vertex 
is given by (cf.\ \eq{tgv})
\be
\tA_{\mu}^{m}(q)A_{\rho}^{r}(k_{1})A_{\sigma}^{s}(k_{2})
\,{\bf :}\qquad\qquad
gf^{mrs}\Bigl(\Gamma_{\mu\rho\sigma}^{F}(q;k_{1},k_{2})
+ (1-\xi_{Q}^{-1})
\Gamma_{\mu\rho\sigma}^{P}(q;k_{1},k_{2})\Bigr)\,\,.
\ee
Thus, choosing the Feynman-like quantum gauge $\xi_{Q} =1$,
neither the quantum gauge field propagator nor the
background-quantum-quantum triple gauge vertex supplies 
factors of 
longitudinal gauge field four-momentum. Thus, in this
particular gauge, at the one-loop level, there are no 
longitudinal factors to trigger
the PT rearrangement (there is ``no pinching'').
In this way, one understands at an heuristic level
how the Feynman rules of the BFM with $\xi_{Q}=1$
give the PT one-loop $n$-point functions directly. 

The relation between the PT and the BFM may be summarised as
follows:

\begin{enumerate}

\item
For $\xi_{Q} = 1$ in the BFM, for
both unbroken and spontaneously broken theories,
at the one-loop level there are no factors of longitudinal internal
gauge field four-momentum to trigger the PT rearrangement.
The BFM at $\xi_{Q}=1$ thus provides the gauge in which one
obtains the PT one-loop $n$-point functions directly
\cite{BFMPT1,BFMPT2,BFMPT3}.

\item
For $\xi_{Q} \neq 1$, the PT algorithm 
can be implemented in the BFM formalism 
to obtain, once again, the PT gauge-independent one-loop
$n$-point functions for a given theory, unbroken or broken
\cite{PTquark}.

\item 
For $\xi_{Q} \neq 1$, the BFM $n$-point
functions display properties (e.g.\ imaginary parts with
unphysical thresholds \cite{PT2pt,PTeweffch}) 
that preclude their use in the cases
described in the introduction.
The PT is thus {\em not}\, just a special
case of the BFM, as claimed in \cite{BFMPT1}.

\item
Beyond one loop, 
the correspondence between the PT $n$-point functions
and those obtained in the BFM with $\xi_{Q} =1$
does {\em not}\, persist \cite{PT2loop,QCD98}.
The correspondence at one loop is thus simply a 
convenient fortuity.

\end{enumerate}

\vspace*{0.3cm}

{\large {\bf 4. Relation to Physical Observables}}

\noindent
It is well known that in QED with fermions $f = e,\mu\ldots$, the
imaginary part of, e.g., the muon contribution to the one-loop
vacuum polarization is directly related to the tree level cross
section for the process $e^{+}e^{-} \rightarrow \mu^{+}\mu^{-}$.
This relation is a result purely of the unitarity of the 
$S$-matrix $S = 1+iT$, expressed in the optical theorem for the
particular case of forward scattering in the process
$e^{+}e^{-} \rightarrow e^{+}e^{-}$:
\be
\im \langle e^{+}e^{-}|T|e^{+}e^{-}\rangle
=
\frac{1}{2}\sum_{i}\int d\Gamma_{i}\,
|\langle i|T|e^{+}e^{-}\rangle|^{2}\,\,.
\label{opth}
\ee
In \eq{opth}, the sum on the r.h.s.\ is over all on-shell physical
states $|i\rangle$ with the quantum numbers of $|e^{+}e^{-}\rangle$;
in each case, the integral is over the available phase space $\Gamma_{i}$.

Consider now the case of the standard electroweak model, in particular
the tree level contribution of the gauge boson pair 
$|i\rangle = |W^{+}W^{-}\rangle$ to the r.h.s.\ of \eq{opth},
i.e.\ the process $e^{+}e^{-} \rightarrow W^{+}W^{-}$. 
The $T$-matrix element for this process is given by
\be
\langle W^{+}W^{-}|T|e^{+}e^{-}\rangle
=
\epsilon_{\mu}^{*}(p_{1})\epsilon_{\nu}^{*}(p_{2})
\,\overline{v}(k_{2})T^{\mu\nu}u(k_{1})\,\,,
\ee

\begin{center}
\begin{picture}(400,105)(0,45)
 
 
\put( 30, 58){\makebox(0,0)[c]{\small $e^{+}(k_{2})$}}
\put( 30,144){\makebox(0,0)[c]{\small $e^{-}(k_{1})$}}
\ArrowLine( 40, 70)( 55,100)
\ArrowLine( 40,130)( 55,100)
\put( 73,115){\makebox(0,0)[c]{\small $\gamma$}}
\Photon( 55,100)( 85,100){4}{3}
\Photon( 85,100)(100, 70){4}{3}
\ArrowLine( 93, 85)( 90.5, 82)
\Photon( 85,100)(100,130){4}{3}
\ArrowLine( 91,115)( 96,115)
\put(115, 58){\makebox(0,0)[c]{\small $W^{+}(p_{2})$}}
\put(115,144){\makebox(0,0)[c]{\small $W^{-}(p_{1})$}}
\put( 70,45){\makebox(0,0)[c]{\small (a)}}
 
\ArrowLine(170, 70)(185,100)
\ArrowLine(170,130)(185,100)
\put(203,115){\makebox(0,0)[c]{\small $Z$}}
\Photon(185,100)(215,100){4}{3}
\Photon(215,100)(230, 70){4}{3}
\ArrowLine(223, 85)(220.5, 82)
\Photon(215,100)(230,130){4}{3}
\ArrowLine(221,115)(226,115)
\put(200,45){\makebox(0,0)[c]{\small (b)}}
 
\ArrowLine(300, 70)(330, 85)
\ArrowLine(300,130)(330,115)
\put(322,100){\makebox(0,0)[c]{\small $\nu_{e}$}}
\Line(330, 85)(330,115)
\Photon(330, 85)(360, 70){-4}{3}
\ArrowLine(345, 77.5)(347, 79)
\Photon(330,115)(360,130){4}{3}
\ArrowLine(345,122.5)(347,121)
\put(330,45){\makebox(0,0)[c]{\small (c)}}
 
\end{picture}
\end{center}
 
\noindent
{\small Fig.~2.
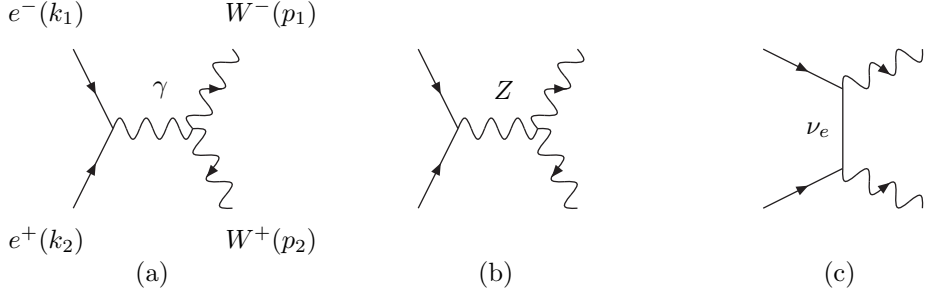
The three diagrams for the process
$e^{+}e^{-} \rightarrow W^{+}W^{-}$ at tree level 
with massless fermions.}
 
\vspace{0.3cm}

\noindent
where, for massless fermions, 
$T^{\mu\nu}$ is given by the sum of the diagrams in Fig.~2,
and $\epsilon_{\mu}^{*}$, $\epsilon_{\nu}^{*}$ are the 
$W^{-}$, $W^{+}$ polarization vectors.
The differential cross section is then given by
\bea
\frac{d\sigma(e^{+}e^{-}\!\rightarrow\! W^{+}W^{-})}{d\Omega}
\!\!&=&\!\!
\frac{1}{64\pi^{2}s}\beta
\frac{1}{4}\!\sum_{e^{\pm}\,\mbox{\scriptsize pols}}
\sum_{W^{\pm}\,\mbox{\scriptsize pols}}
|\langle W^{+}W^{-}|T|e^{+}e^{-}\rangle|^{2} \\
\!\!&=&\!\!
\frac{1}{64\pi^{2}s}\beta
\frac{1}{4}\!\sum_{e^{\pm}\,\mbox{\scriptsize pols}}
(\overline{v}T_{\mu'\!\nu'}u)^{*}
\biggl(g^{\mu'\!\mu} - \frac{p_{1}^{\mu'\!}p_{1}^{\mu}}{M_{W}^{2}}\biggr)
\biggl(g^{\nu'\!\nu} - \frac{p_{2}^{\nu'\!}p_{2}^{\nu}}{M_{W}^{2}}\biggr)
(\overline{v}T_{\mu\nu}u)  \qquad
\label{wwpols}
\eea
where $s = (p_{1}+p_{2})^{2}$ and $\beta = \sqrt{1 - 4M_{W}^{2}/s}$. 
In \eq{wwpols}, the sums over the $W^{\pm}$ polarizations have
been written explicitly in terms of the corresponding four-momenta.

The differential cross section in \eq{wwpols}
may be decomposed in two distinct ways:
\be
\frac{d\sigma(e^{+}e^{-}\rightarrow W^{+}W^{-})}{d\Omega}
=
\underbrace{
\sum_{i,j}\frac{d\sigma_{ij}(e^{+}e^{-}\rightarrow W^{+}W^{-})}{d\Omega}
}_{\mbox{\scriptsize conventional decomposition}}
=
\underbrace{
\sum_{i,j}\frac{d\hat{\sigma}_{ij}(e^{+}e^{-}\rightarrow W^{+}W^{-})}{d\Omega}
}_{\mbox{\scriptsize PT decomposition}}\,\,,
\ee
where $i,j = \gamma, Z, \nu$. In the conventional decomposition \cite{buras},
the $d\sigma_{ij}/d\Omega$ correspond directly to the diagrams in 
Fig.~2. For example, $d\sigma_{\gamma\gamma}/d\Omega$ is the
contribution of the diagram in Fig.~2a, while 
$d\sigma_{Z\nu}/d\Omega$ is the $Z\nu$ interference term. As is well
known, these contributions individually diverge at high $s$,
and so violate unitarity. Nevertheless, their sum is well-behaved.
However, the bad high-energy behaviour of the 
$d\sigma_{ij}/d\Omega$ precludes the use of the optical theorem
(\ref{opth}) to interpret {\em individually}\, the components
$\{\sigma_{\gamma\gamma}, \sigma_{\gamma Z}, \sigma_{ZZ}\}$,
$\{\sigma_{\gamma\nu}, \sigma_{Z\nu}\}$ and $\{\sigma_{\nu\nu}\}$
of the tree level cross section for
$e^{+}e^{-}\rightarrow W^{+}W^{-}$ in terms of the imaginary
parts of renormalizable one-loop self-energy-, vertex- and box-like
$W^{+}W^{-}$ contributions, respectively, to the process 
$e^{+}e^{-}\rightarrow e^{+}e^{-}$.

In the PT decomposition \cite{PTeweffch}, one first contracts 
into $(\overline{v}T_{\mu\nu}u)$, $(\overline{v}T_{\mu'\!\nu'}u)^{*}$
the longitudinal four-momenta appearing in the $W^{\pm}$ 
polarization sums in \eq{wwpols}, triggering the elementary
Ward identity obeyed by $T_{\mu\nu}$. 
This Ward identity (i) enforces the exact cancellation among
contributions from diagrams with distinct $s$- and $t$-channel
propagator structure and
which individually diverge at high energy, and
(ii) specifies the non-vanishing contributions, of purely non-abelian
origin, which remain after this cancellation.
The components $d\hat{\sigma}_{ij}/d\Omega$ in the PT
decomposition are then defined by the 
$i,j = \gamma,Z,\nu$ propagator structure obtained {\em after}\,
the systematic and exhaustive implentation
of this Ward identity.
It is found \cite{PTeweffch}
that the components  $d\hat{\sigma}_{ij}/d\Omega$
are {\em individually}\, well-behaved at high energy.
In particular, the three self-energy-like components
are proportional at high energy to the $W$ contribution
to the corresponding electroweak $\beta$ functions
(and so are negative).

\begin{center}

\begin{picture}(400,80)(0,10)
 
 
\ArrowLine(150, 75)(200, 50)
\ArrowLine(150, 25)(200, 50)
\Photon(200,50)(240,70){-4}{5}
\ArrowLine(240, 70)(260, 80)
\ArrowLine(240, 70)(260, 60)
\Photon(200,50)(240,30){ 4}{5}
\ArrowLine(240, 30)(260, 40)
\ArrowLine(240, 30)(260, 20)
\GCirc(200, 50){13}{0.8}

\put(140, 75){\makebox(0,0)[c]{\small $e^{-}$}}
\put(140, 25){\makebox(0,0)[c]{\small $e^{+}$}}

\put(220, 80){\makebox(0,0)[c]{\small $W^{-}(p_{1})$}}
\put(220, 20){\makebox(0,0)[c]{\small $W^{+}(p_{2})$}}

\put(270, 80){\makebox(0,0)[c]{\small $f_{1}$}}
\put(270, 60){\makebox(0,0)[c]{\small $f_{2}$}}
\put(270, 40){\makebox(0,0)[c]{\small $f_{3}$}}
\put(270, 20){\makebox(0,0)[c]{\small $f_{4}$}}

\end{picture}
\end{center}

\noindent
{\small Fig.~3.
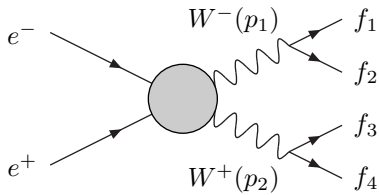
The schematic representation of the LEP2 process 
$e^{+}e^{-} \rightarrow W^{+}W^{-} \rightarrow  \mbox{four fermions}$. }

\vspace{0.3cm}

One may then define, by {\em direct analogy}\, with QED,
the imaginary parts of one-loop
$W^{+}W^{-}$ (including associated would-be Goldstone boson and ghost)
contributions to self-energy-like functions
$\hat{\Sigma}_{ij}$ directly in terms of the corresponding
self-energy-like components 
of the tree level $e^{+}e^{-} \rightarrow W^{+}W^{-}$ cross section:
\bea
e^{2}\frac{1}{s^{2}}
\,\im\hat{\Sigma}_{\gamma\gamma}^{(WW)}(s)
&=&
\hat{\sigma}_{\gamma\gamma}(e^{+}e^{-}\rightarrow W^{+}W^{-}) \\
2e^{2}\frac{a}{s_{w}c_{w}}\, \frac{1}{s(s-M_{Z}^{2})}
\,\im\hat{\Sigma}_{\gamma Z}^{(WW)}(s)
&=&
\hat{\sigma}_{\gamma Z}(e^{+}e^{-}\rightarrow W^{+}W^{-}) \\
e^{2}\frac{(a^{2}+b^{2})}{s_{w}^{2}c_{w}^{2}}\,\frac{1}{(s-M_{Z}^{2})^{2}}
\,\im\hat{\Sigma}_{ZZ}^{(WW)}(s)
&=&
\hat{\sigma}_{ZZ}(e^{+}e^{-}\rightarrow W^{+}W^{-})\,\,,
\eea
where $s_{w}$ $(c_{w})$ is the sine (cosine) of the weak mixing angle,
$a = \frac{1}{4} - s_{w}^{2}$, $b=\frac{1}{4}$.
Given the imaginary parts of the self-energy functions, the full
renormalized functions may then be constructed from twice-subtracted
dispersion relations. 
It is found that the resulting functions are precisely those obtained
at one loop in the PT \cite{PTeweffch}.

The significance of this result is that the individual differential
cross section components $d\hat{\sigma}_{ij}/d\Omega$
can in principle be projected out from the full
differential cross section \cite{PTeweffch}. 
In this way, the PT self-energy
functions are directly related to physical observables.

\vspace*{0.3cm}

{\large {\bf 5. Example Application: Unstable Particles}}

\noindent
Perhaps the most important application of the PT is to the physics
of unstable particles \cite{PT2pt,higgs}. 
For example, consider the LEP2 process 
$e^{+}e^{-} \rightarrow W^{+}W^{-} \rightarrow \mbox{4 fermions}$
involving intermediate $W$ pair production, 
as shown in Fig.~3.
At tree level, the  $W$ propagators are singular at
$p_{i}^{2}  = M_{W}^{2}$, $i=1,2$. 
Clearly, in order to evalute amplitudes for such processes
at arbitrary values of the kinematic parameters,
it is essential to regulate such singularities.
This involves the introduction in some way of a finite
decay width.

Originally, two approaches were popular. In the ``fixed width''
approach, one makes the systematic replacement
$(q^{2}-M^{2})^{-1} \rightarrow (q^{2}-M^{2} + iM\Gamma)^{-1}$
for the propagators,
where $\Gamma$ is the physical width of the particle of mass $M$. 
However, the introduction of a non-zero width for space-like $q^{2}$
is clearly unphysical, and leads to violations of unitarity.
In the ``running width'' approach, the introduction of a $q^{2}$-dependent
width $\Gamma(q^{2})$ enables unitarity to be restored. 
In general, however, both of these approaches lead to violations
of gauge invariance. Although in favourable cases, such as at LEP1,
the resulting effects can be shown to be small, in other cases,
such as those involving longitudinal gauge bosons at high energies,
the loss of gauge invariance can lead to catastrophically large effects, 
and hence useless results.

The fundamental problem is that the decay width of an
unstable particle arises from an {\em infinite subset}\, of diagrams,
i.e.\ from the Dyson summation of the 1PI self-energy
$\Sigma(q^{2})$. This leads to the replacement
\be
\frac{1}{q^{2} - M^{2}}
\longrightarrow
\frac{1}{q^{2} - M^{2} + \re \Sigma(q^{2}) + i\,\im \Sigma(q^{2})}\,\,.
\ee
At $q^{2} = M_{p}^{2}$, where $M_{p}$ is the (complex)
pole mass, defined from
$M_{p}^{2} - M^{2} + \Sigma(M_{p}^{2}) = 0$,
the function $\Sigma(q^{2})$ is gauge-independent.
In particular, at the one-loop level,
the lowest order width $\Gamma^{(1)}$ 
is given by $M\Gamma^{(1)} = \im \Sigma^{(1)}(M^{2})$.
At $q^{2} \neq M_{p}^{2}$, however,
the conventional self-energy $\Sigma(q^{2})$ 
is in general gauge-dependent. 
For a given process, 
summing the infinite subset of contributions involved
in the Dyson series while including the remaining contributions
only to some finite order then in general
leads to two  problems (see e.g.\ \cite{argyres}).
First, the delicate order-by-order gauge cancellation mechanism
is distorted, so that gauge independence is lost.
And second, the Ward identities are violated, so that gauge invariance
is also lost.

The most satisfactory approach to these problems is to
attempt to identify the infinite subset of contributions which
encode the physics of the unstable particles.
In particular, the observation that, at lowest order,
the gauge bosons of the standard model decay exclusively
into fermions has led to the ``fermion loop scheme'' \cite{FLS}.
In this approach, for a given tree level process involving
unstable gauge bosons, e.g.\ Fig.~3, one includes
{\em all}\, possible fermionic one loop corrections,
with those contributing to the self-energy summed in Dyson series.
This infinite subset of corrections is both gauge-independent,
since the individual fermion loop corrections are gauge-independent,
and gauge-invariant, since the ensemble of fermion loop corrections,
including the Dyson-summed self-energies, exactly
satisfy tree-level-like Ward identities.

However, it is known that bosonic corrections too can produce large effects,
especially at high energies. In order to account for bosonic effects, 
the BFM has been advocated as an appropriate framework \cite{dennditt}.
However, as already stated, although
the ensemble of BFM $n$-point functions are gauge-invariant, they
remain individually gauge- i.e.\ $\xi_{Q}$-dependent. In particular,
for $\xi_{Q} \neq 1$ their imaginary parts have unphysical thresholds
\cite{PT2pt,PTeweffch}.
Thus, the naive Dyson summation of the BFM self-energies leads to 
gauge-dependent matrix elements. 

By contrast, the PT provides a general framework for the 
treatment of unstable particles at the one-loop level 
in which not only gauge invariance {\em and}\, gauge independence
are maintained, 
but also a range of further requisite field-theoretic
properties are satisfied. 
Detailed discussions of these issues are given in \cite{PT2pt}
for gauge bosons and \cite{higgs} for the Higgs boson.

\vspace*{0.3cm}

{\large {\bf 6. Progress Beyond One Loop}}

\noindent
Most recently, work on the PT has concentrated on the attempt
to extend the algorithm beyond one loop \cite{PT2loop,QCD98}.
{}From a theoretical point of view, 
this extension is clearly essential if it is to be shown that the PT
is more than just an artefact of the one-loop approximation.
{}From a phenomenological point of view,
many of the applications of the PT increasingly
demand accuracy beyond the one-loop level.
There are two basic problems:

\begin{enumerate}

\item
{\em How to deal consistently with triple gauge vertices all three legs
of which are associated with gauge fields propagating in loops?}\,
In the PT at the one-loop level, 
the factors of longitudinal four-momentum associated with gauge fields 
propagating in loops originate from tree level gauge 
field propagators and triple gauge vertices. In particular, 
the triple gauge vertices which occur in one-loop diagrams
always have one external leg $A_{\mu}^{a}(q)$ and
two internal legs $A_{\rho}^{r}(k_{1})$, $A_{\sigma}^{s}(k_{2})$.
It is then possible to decompose such vertices so as to isolate
unambiguously the longitudinal factors $k_{1\rho}$, $k_{2\sigma}$
associated with the internal gauge fields.
Beyond one loop, however, there occur triple gauge vertices for
which all three legs are internal.
It is thus unclear how to decompose such vertices
in order to identify the associated longitudinal factors
which then trigger the PT rearrangement.

\item
{\em How to deal consistently with the ``induced''
factors of longitudinal internal gauge field four-momentum 
which originate from internal loop corrections?}
Beyond the one-loop 

\begin{center}

\begin{picture}(440,60)(25,340)
 
 
\ArrowLine(  0,400)( 14,400)
\GCirc( 25,400){11}{0.8}
\put( 25,401){\makebox(0,0)[c]{\Large{\bf 2}}}
\ArrowLine( 36,400)( 50,400)
\put( 25,375){\makebox(0,0)[c]{\small (a)}}
 
\put( 57.5,400){\makebox(0,0)[c]{$=$}}
 
\Line( 65,400)(115,400)
\GlueArc( 90,400)(15,  0,180){1.8}{8}
\GCirc( 90,400){6}{0.8}
\put( 91,401){\makebox(0,0)[c]{{\bf 1}}}
\put( 90,375){\makebox(0,0)[c]{\small (b)}}

\put(122.5,400){\makebox(0,0)[c]{$+$}}
 
\ArrowLine(130,400)(180,400)
\GlueArc(155,400)(15,  0, 90){1.8}{4}
\GlueArc(155,400)(15, 90,180){1.8}{4}
\GCirc(155,415){6}{0.8}
\put(156,416){\makebox(0,0)[c]{{\bf 1}}}
\put(155,375){\makebox(0,0)[c]{\small (c)}}
 
\put(187.5,400){\makebox(0,0)[c]{$+$}}
 
\ArrowLine(195,400)(245,400)
\GlueArc(220,400)(15,  0,180){1.8}{8}
\GCirc(235,400){6}{0.8}
\put(236,401){\makebox(0,0)[c]{{\bf 1}}}
\put(220,375){\makebox(0,0)[c]{\small (d)}}
 
\put(252.5,400){\makebox(0,0)[c]{$+$}}
 
\ArrowLine(260,400)(310,400)
\GlueArc(285,400)(15,  0,180){1.8}{8}
\GCirc(270,400){6}{0.8}
\put(271,401){\makebox(0,0)[c]{{\bf 1}}}
\put(285,375){\makebox(0,0)[c]{\small (e)}}
 
\put(317.5,400){\makebox(0,0)[c]{$-$}}
 
\ArrowLine(325,400)(375,400)
\GlueArc(350,400)(15,  0,180){1.8}{8}
\GBox(345,395)(355,420){0.8}
\put(351,408){\makebox(0,0)[c]{{\bf 0}}}
\put(350,375){\makebox(0,0)[c]{\small (f)}}

\put(382.5,400){\makebox(0,0)[c]{$+$}}
 
\ArrowLine(390,400)(415,400)
\ArrowLine(415,400)(440,400)
\GCirc(415,400){2.5}{0}
\put(415,375){\makebox(0,0)[c]{\small(g)}}
 
\put(  0,353){\makebox(0,0)[l]{\small Fig.~4.
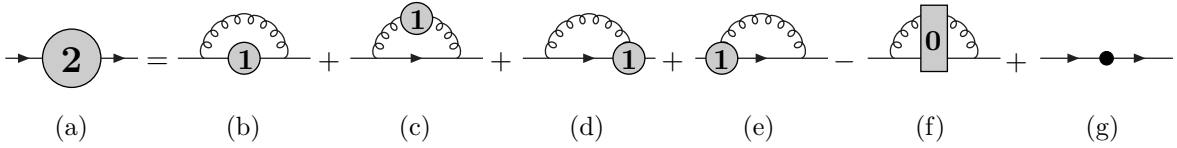
The$\!$ skeleton $\!$expansion$\!$ representation $\!$of the
$\!$renormalized $\!$two-loop $\!$quark $\!$self-energy. The$\!$ box
}}
 
\put(  0,340){\makebox(0,0)[l]{\small
in (f) represents the 
lowest order contribution to the Bethe-Salpeter quark-gluon 
scattering kernel.
}}

\end{picture}
\end{center}

\vspace{0.3cm}

\noindent
level, in addition to
the factors of longitudinal internal gauge field four-momentum
from tree level gauge field propagators and triple gauge vertices,
there occur further
such factors originating from the invariant tensor structure of
internal loop corrections. A simple example are the longitudinal
factors occurring in the transverse structure
of the gluon self-energy in QCD.
It is thus unclear whether or not such ``induced'' factors
should also be used to trigger the PT rearrangement;
and, if so, how this may be done consistently.

\end{enumerate}
At the two-loop level, the two problems are uncoupled and 
so can be investigated separately and successively: one first
of all has to solve (1) in order to identify the one-loop
internal corrections needed before one can address (2).

The approach to problem (1) taken in \cite{PT2loop} is based on the
assumption that there exists a {\em skeleton expansion}\, for QCD 
\cite{QCDskeleton}
analogous to that of QED \cite{bjdr}. In the case of the quark self-energy,
such a diagrammatic
representation is easy to construct, and involves the
concept of a Bethe-Salpeter scattering kernel. This representation follows
from the Schwinger-Dyson equations for QCD \cite{SDreview}.
Expanded out at
the two-loop level, this representation is as shown in Fig.~4,
with the lowest order kernel as shown in Fig.~5.
In Figs.~4 and 5, the one-loop internal corrections and
the tree level propagators are the conventional
gauge-dependent functions.

It was shown in \cite{PT2loop} that, demanding that the renormalized
one-loop internal corrections in Figs.~4b--4e be the PT
gauge-independent one-loop functions, 
the corresponding PT kernel term which {\em automatically}\,
results from the necessary rearrangement is as shown in Fig.~6. 
Thus, the PT kernel is identical to the conventional kernel
in the Feynman gauge, except that the triple gauge vertex in Fig.~6c 
involves only the component $\Gamma^{F}$
(cf.\ \eq{tgv}), with orientation as indicated.
Crucially, the PT kernel thus
supplies no longitudinal factors associated with the external gluon legs,
and so, when embedded in Fig.~4f,
cannot trigger any further PT rearrangement.
In this way, the two-loop PT rearrangement for the quark self-energy
is self-consistent.

It was explicitly shown \cite{PT2loop} that this PT two-loop
quark self-energy is gauge-independent at all momenta,
does not shift the position of the propagator pole, and
is multiplicatively renomalizable by local counterterms.
Furthermore, it was shown by this example that
the general correspondence between the PT $n$-point functions
and those obtained in the BFM at $\xi_{Q}=1$ does not
persist beyond one loop.

\begin{center}

\begin{picture}(440,105)(0,330)
 
 
\ArrowLine(  0,400)( 20,400)
\ArrowLine( 30,400)( 50,400)
\Gluon(  0,415)( 50,415){1.8}{11}
\GBox( 20,395)( 30,421){0.8}
\put( 26,408){\makebox(0,0)[c]{{\bf 0}}}
\put( 25,380){\makebox(0,0)[c]{\small (a)}}
 
\put( 57.5,408){\makebox(0,0)[c]{$=$}}
 
\ArrowLine( 65,400)(115,400)
\Gluon( 75,400)(115,415){1.8}{10}
\Gluon( 65,415)( 85,408){1.8}{4}
\Gluon( 95,404)(105,400){1.8}{2}
\put( 90,380){\makebox(0,0)[c]{\small (b)}}

\put(122.5,408){\makebox(0,0)[c]{$+$}}
 
\ArrowLine(130,400)(155,400)
\ArrowLine(155,400)(180,400)
\Gluon(155,400)(155,415){1.8}{3}
\Gluon(130,415)(155,415){1.8}{5}
\Gluon(155,415)(180,415){1.8}{5}

\put(155,380){\makebox(0,0)[c]{\small (c)}}
 
 
\ArrowLine(260,400)(280,400)
\ArrowLine(290,400)(310,400)
\Gluon(260,415)(310,415){1.8}{11}
\GBox(280,395)(290,421){0.8}
\put(286,415){\makebox(0,0)[c]{\tiny {\bf PT}}}
\put(286,404){\makebox(0,0)[c]{{\bf 0}}}
\put(285,380){\makebox(0,0)[c]{\small (a)}}
 
\put(317.5,408){\makebox(0,0)[c]{$=$}}
 
\ArrowLine(325,400)(375,400)
\Gluon(335,400)(375,415){1.8}{10}
\Gluon(325,415)(345,408){1.8}{4}
\Gluon(355,404)(365,400){1.8}{2}
\put(350,380){\makebox(0,0)[c]{\small (b)}}

\put(382.5,408){\makebox(0,0)[c]{$+$}}
 
\Line(400,422)(430,422)
\ArrowLine(390,400)(415,400)
\ArrowLine(415,400)(440,400)
\Gluon(415,400)(415,415){1.8}{3}
\Gluon(390,415)(415,415){1.8}{5}
\Gluon(415,415)(440,415){1.8}{5}
\put(415,380){\makebox(0,0)[c]{\small (c)}}

\put(  0,358){\makebox(0,0)[l]{\small Fig.~5.
The contributions to the lowest order
}}
 
\put(  0,345){\makebox(0,0)[l]{\small
conventional Bethe-Salpeter quark-gluon 
scat-}}
 
\put(  0,332){\makebox(0,0)[l]{\small tering kernel.
}}
 
\put(250,358){\makebox(0,0)[l]{\small Fig.~6.
The contributions to the PT kernel.
}}
 
\put(250,345){\makebox(0,0)[l]{\small The propagator and
vertex in (c) are $D^{F}$ 
}}

\put(250,332){\makebox(0,0)[l]{\small and
$\Gamma^{F}$ (line indicates orientation of $\Gamma^{F}$).
}}

\end{picture}
\end{center}

The quark self-energy is the first non-trivial $n$-point function
to be tackled beyond one loop in the PT approach. 
Despite the progress which has been made,
the general solution of the two problems described above 
appears still to represent a formidable challenge in field theory.
For a discussion of problem (2), see \cite{QCD98}.

\vspace*{0.3cm}

{\large {\bf 7. Conclusions}}

\noindent
The PT provides an approach to perturbation theory
at the one-loop level in which the one-loop $n$-point functions
are both {\em gauge-independent},
i.e.\ they are individually entirely independent of the 
particular gauge-fixing procedure used, and 
{\em gauge-invariant}, 
i.e.\ they satisfy simple tree-level-like Ward identities
associated with the gauge symmetry of the classical lagrangian.
In addition to these two basic properties,
the PT $n$-point functions display a wide range of further
desirable field-theoretic properties. 
The PT thus provides a general perturbative framework
which enables one to deal with
a broad range of applications in which one is
forced to go beyond the strictly order-by-order computation of
$S$-matrix elements, or to consider amplitudes for explicitly
off-shell processes. 

Despite the appeal of the PT approach, there remain, however, 
two basic criticisms which need to be answered. First, it must now
be shown in general how the PT can be extended beyond the one
loop approximation. And second, it is now highly desirable to have
a more formal understanding of the PT, beyond diagrammatics.
Thus, there remains much work to be done if the PT approach is to 
fulfill its promise.

\vspace{0.3cm}

It is a pleasure to thank Joannis Papavassiliou for many 
illuminating discussions.

\end{document}